\documentclass[aps,prl,twocolumn]{revtex4}
\usepackage{graphicx}
\usepackage{natbib}

\def\aas{{ Astron. \& Astrophys.\ Suppl. \ }}
\def\apj{{Astrophys. \ J. \ }}

\def\np{{ Nucl.\ Phys.}}
\def\plb{{ Phys.\ Lett.\ B \ }}

\def\prl{{ Phys.\ Rev.\ Lett.\ }}

\newcommand{\lsim}{\,\lower2truept\hbox{${<\atop\hbox{\raise4truept\hbox{$\sim$}}}$}\,}
\newcommand{\gsim}{\,\lower2truept\hbox{${>\atop\hbox{\raise4truept\hbox{$\sim$}}}$}\,}

\newcommand{\be}{\begin{equation}}
\newcommand{\ee}{\end{equation}}
\newcommand{\bea}{\begin{eqnarray}}
\newcommand{\eea}{\end{eqnarray}}

\begin{document}

\title{Approaching Lambda without fine-tuning}
\author{Sabino Matarrese$^{1,2} \footnote{sabino.matarrese@pd.infn.it}$, 
Carlo Baccigalupi$^{3,4} \footnote{bacci@sissa.it}$, 
Francesca Perrotta$^{3,4} \footnote{perrotta@sissa.it}$}
\affiliation{$^1$ Dipartimento di Fisica Galileo Galilei, Universit\'a 
di Padova, \\\
$^2$ INFN, Sezione di Padova, via Marzolo 8, I-35131, Padova, Italy, \\
$^3$ SISSA/ISAS, via Beirut 4, 34014 Trieste, Italy, \\
$^4$ INFN, Sezione di Trieste, via Valerio 2, I-34127 Trieste, Italy} 

\begin{abstract}

We address the fine-tuning problem of dark energy cosmologies which arises 
when the dark energy density needs to initially lie 
in a narrow range in order for its present value to be 
consistent with observations. 
As recently noticed, this problem becomes particularly severe in 
canonical Quintessence scenarios, when trying to reproduce  
the behavior of a cosmological constant, i.e. when the dark 
energy equation of state $w_Q$ approaches $-1$: these models may be 
reconciled with a large basin of attraction only by requiring a rapid
evolution of $w_Q$ at low reshifts, which is in conflict with the most
recent estimates from type Ia Supernovae discovered by Hubble Space 
Telescope. 
Next, we focus on scalar-tensor theories of gravity, discussing the 
implications of a coupling between the Quintessence scalar field and 
the Ricci scalar (``Extended Quintessence''). We show that, 
even if the equation of state today is very close to $-1$, by virtue of the 
scalar-tensor coupling the quintessence trajectories still possess 
the attractive feature which allows to reach the 
present level of cosmic acceleration starting by a set of initial 
conditions which covers tens of orders of magnitude; this 
effect, entirely of gravitational origin, represents a new 
important consequence of the possible coupling between dark energy and
gravity. 
We illustrate this effect in typical Extended Quintessence scenarios. 
\end{abstract}

\maketitle

Since the observations of distant type Ia Supernovae suggested that 
the Universe expansion is accelerating \cite{SN1aR,SN1aP}, a great deal 
of cosmological models has been proposed in order to describe possible 
mechanisms to speed up the expansion rate. 
Most of the models in the literature focus on ``Quintessence''  
cosmologies, where a classical, minimally-coupled 
scalar field evolves along a shallow potential, while its energy density and 
pressure combine to produce a negative equation of state, thus making 
the field act as a repulsive force.
The fine tuning problem of many Quintessence models, i.e. the need
to set the field initial conditions in a very tiny range in 
order to get the observed energy density and equation of state today 
(as distinguished from the ``coincidence problem'', i.e. the similarity 
between the dark and the critical energy densities today), 
 motivated the search for models where 
the equations of motion admits attractor solutions \cite{CDS}-\cite{Wetterich}.
The main property of attractor solutions is that a very wide range 
of initial conditions rapidly converge to a common evolutionary track.
In particular, ``tracking'' scalar fields will eventually evolve into a 
``tracking solution'' in the background-dominated era, where they 
have almost constant equation of state $w_Q$ lying between $-1$ and the 
background equation of state \cite{SWZ,LS}. 
Attractor solutions can be obtained from different potentials; 
the most popular being the exponential potential \cite{Wetterich}, the 
inverse 
power-law potential \cite{RP}, and the SUGRA model \cite{BM}, where the 
exponential function is modulated by an inverse power law.
Despite the appeal of these models, originally introduced to overcome 
the fine-tuning problem while allowing for a negative equation of 
state, it has been recently pointed out in \cite{Bludman} 
that they are affected by a serious drawback when analyzed in 
the post-tracking regime.  
The most recent analysis, combining Cosmic Microwave Background (CMB) 
observations, Large-Scale Structure (LSS) data, Hubble parameter estimation 
and distant type Ia Supernovae, give, for a fiducial model with constant 
Quintessence equation of state, $w_Q = -0.98 \pm 0.12 $
 \cite{Spergel}, while other authors give 
$w_Q=-0.91^{+0.13}_{-0.15}$  \cite{WT}. A completely orthogonal 
information comes from the age of globular clusters, giving $w_Q < -0.8$
($68 \% \ $confidence level) \cite{Jimenez}. 
The problem with tracking Quintessence is that, once we fix the present 
Quintessence energy density to a value consistent with its estimate 
in a flat universe, $\Omega_Q = 0.73 \pm 0.04$ 
\cite{Spergel}, the observational constraints on $w_Q$ allow only 
``crawling'' quintessence, or potentials with large current curvature. 
The tracking solutions are, indeed, defined in the background dominated 
epoch, where the background is either radiation or non-relativistic matter; 
in the present Quintessence-dominated era, the scalar field has already 
passed the tracking phase. When tracking models are analyzed in detail, 
including the post-tracking behavior of the field, it can be 
shown that ``good trackers'', i.e.  attractors having a large basin 
of attraction, end up in the present Quintessence dominated era with 
a value of the equation of state which, being too different from 
$-1$, is ruled out by the observational constraints. 
More plausible values of $w_Q$ today can be obtained by flattening the 
potential in which the field evolves (``crawling'' Quintessence), but 
this requires to shrink the basin of attraction (``poor trackers''), 
making the Universe too sensitive to the Dark Energy (DE) initial 
conditions. This shrinking becomes increasingly dramatic the more the 
present $w_Q$  approaches the cosmological constant value $-1$, in which 
case one has to exactly tune the initial value of the primordial 
Quintessence energy density to the same value it has today. \\

Good trackers may actually be reconciled with observations only if the 
potential curvature (and, therefore, $w_Q$) is rapidly varying 
at redshifts $z \lsim 0.5$, but this appears to further exacerbate the 
coincidence problem, in that we require the energy density to be 
small {\it and} rapidly changing at recent times. 
Even accepting an anthropic explanation to this coincidence, another 
potential problem for good trackers with rapid current evolution of 
the field may arise from observational constraints on $\dot 
w_Q \equiv {dw_Q / dz}$; indeed, the latest analysis of type Ia 
Supernovae with Hubble Space Telescope \cite{Riess2}, combined with 
independent constraints from CMB and LSS data, gives 
$\dot w_Q=0.6 \pm 0.5$, for a flat universe, ruling out the possibility 
of a rapidly-changing equation of state of DE. 
As a consequence, it is clear that 
values of $w_Q$ very close to $-1$ would seriously challenge the 
whole class of canonical Quintessence models. \\

This puzzling problem is however alleviated if we extend
the class of DE models, allowing for a non-minimal coupling 
between the Quintessence scalar field and the 
Ricci scalar \cite{Dolgov}-\cite{CMM}. We will show 
that in Extended Quintessence models of Dark Energy \cite{PBM,BMP}, 
the basin of attraction is enlarged by the so-called R-boost, a purely 
gravitational effect due to the coupling. As a result, non-minimally 
coupled models of Quintessence can preserve 
the appealing feature of attractors which is lost in minimally-coupled 
models, because they can approach the value $w_Q = -1$, while still 
allowing a large range of initial energy density of the field. \\

As discussed in detail in \cite{Bludman}, if we exclude 
potentials whose curvature increases rapidly near the present 
epoch (which would further enhance the coincidence problem), canonical 
Quintessence scenarios require flat potentials in order to reproduce 
an equation of  state close to $-1$; in this case, the trajectories of 
the field are almost flat, so that a real tracking never occurs 
(``crawling'' Quintessence). The range of initial values of the scalar 
field energy density is increasingly narrow as we approach the limiting 
case $w_Q=-1$; the cosmological constant case is realized only 
starting from a single value which is exactly tuned to the present 
one ($\rho_Q \sim 26 \ {\rm meV}^4$ for a flat model with $h=0.7$).
This can be easily understood recalling that, in the tracking regime, 
the value of the DE equation of state is strictly related to 
the shape of the potential; generally, the closer $w_Q$ is 
to $-1$, the flatter the potential needs to be. In practice, the 
efficiency of tracking is lost when requiring such low values of the 
equation of state.\\

\begin{figure} 
\begin{center}
\includegraphics[width=3.0in, height=1.9in]{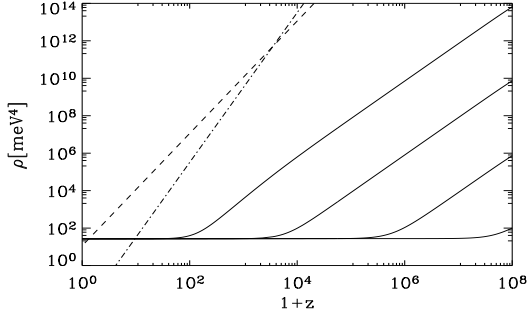}
\caption{Solid curves: energy density of the tracking Extended 
Quintessence with present equation of state $w_Q=-0.999$, with 
different values of the initial scalar field energy density.
Dashed curve: matter energy density. Dot-dashed: radiation. 
}
\label{Figura1}
\end{center}
\end{figure}

\begin{figure}
\begin{center}
\includegraphics[width=3.0in,height=1.9in]{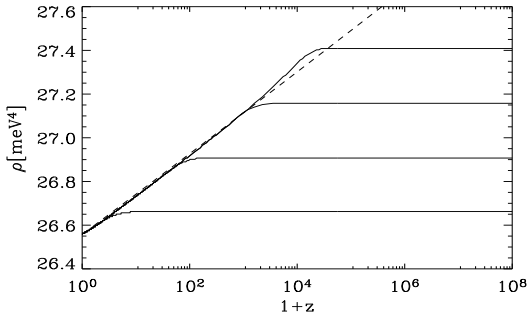}
\caption{Tracking Quintessence energy density in minimally-coupled 
models with inverse power-law potential and 
$w_Q=-0.999$ today.
The curves correspond to different initial values of the field,
which are the same as in Fig. \ref{Figura1}; 
the dashed curve is the attractor solution. Here 
the $y-$axis is plotted on linear scale.}
\label{Figura2}
\end{center}
\end{figure}

However, this shrinking of the basin of attraction is peculiar of 
minimally coupled Quintessence fields: the Klein-Gordon equation 
rules the dynamics of the field, which freezes during most of the 
Universe history if the potential is nearly flat. The situation is 
dramatically different in scalar-tensor theories, where the Ricci 
scalar $R$ in the gravitational sector of the Lagrangian of General 
Relativity is replaced by the product of $R$ with a function $F(\phi)$. 
The most important effect of the non-minimal coupling is to 
enhance the dynamics of the field, while keeping the potential flat. 
As discussed in \cite{BMP}, the coupling adds a new source term in the 
Klein-Gordon equation: 
\begin{equation}
{\phi}''+2 {\cal{H}} {\phi}'= {a^2 \over 2} 
F_{,\phi}R-a^2V_{,\phi} \ \ \ ;
\label{KG}
\end{equation}
in Eq. (\ref{KG}), primes indicate differentiation with respect to 
the conformal time, and ${\cal{H}}\equiv {a' / {a}}$. The new 
term proportional to the Ricci scalar on the R.H.S. adds up to the one 
produced by the ``true'' potential 
of the field, $V(\phi)$, thus generating an effective potential 
which is different from the corresponding minimal-coupling one; the 
difference is especially relevant at high redshifts, when the coupling 
term dominates over the true potential. 
Deep in the radiation era, the quantity $a^2R$ diverges as $a^{-1}$  
imprinting a 
boost of energy to the scalar field, an effect we named ``$R$-boost'' 
\cite{BMP}. This gravitational effect has a deep connection with 
particle physics, because the onset of the $R$-boost, in the  
radiation-dominated epoch, occurs as soon as the first particle 
species of the cosmic plasma becomes non-relativistic \cite{BMP}. \\ 
In typical non-minimal coupling models, where the coupling term is 
proportional to $\phi^2R$,
the amount of $R$-boost depends on the initial value of the scalar 
field. Starting from some initial value of the scalar field energy 
density, the field acquires a new potential energy which will be 
rapidly converted into kinetic energy; in a very short time, the field 
accelerates until the friction caused to the Hubble drag term 
in Eq. (\ref{KG}), becomes comparable with the coupling term. 
After that, the scalar field slowly rolls with a kinetic energy 
scaling as $a^{-2}$ ($a^{-3}$ in the matter epoch), until the true 
potential V becomes important (typically, in the matter dominated 
era), and the evolution occurs 
along the corresponding attractor trajectory: depending on the 
value  $w_Q$ today, there can be a period of freezing with $w_Q \sim 
-1$, followed by tracking, or a ``crawling'' Quintessence without ever 
really ``tracking'' if the present $w_Q$ approaches $-1$.
For inverse power-law potentials, $V(\phi) \propto \phi^{-\alpha}$, 
the equation of state in the tracking regime is 
${w_Q}_{track}=-2/(\alpha+2)$: in order to approach the cosmological 
constant equation of state, the potential is forced to be extremely flat. 
What we want to outline here is that, even in the limiting case 
$w_Q\rightarrow -1$, the $R$-boost enlarges the allowed range of 
initial energy densities ${\Omega_Q}_{init}$, which can cover several 
orders of magnitude, as opposite to the minimally-coupled Quintessence
case, where a flat potential implies early freezing and 
narrow basin of attraction.     
To give a practical example, let us consider the non-minimal coupling 
described in \cite{PBM,BMP}, with $F(\phi)=\xi\phi^2+const.$; 
$\xi$ is the coupling parameter, related to the Jordan-Brans-Dicke 
parameter, $\omega_{JBD} \equiv \left({F 
/{F_{,\phi}}^2}\right)_{today}$. 
We assumed an inverse power-law potential for the 
field, and $\omega_{JBD}=4000$ \cite{OmegaJBD}; though recent 
experiments seem to converge towards larger values of $\omega_{JBD}$ 
\cite{Bertotti}, 
this would only anticipate the onset of the $R$-boost, without affecting 
the substance of our results. We performed 
a numerical integration of the background equations for a flat 
universe, requiring $w_Q=-0.999$, $h=0.7$ and 
$\Omega_Q=0.7$ today. Such an equation of state is so close to $-1$ 
that the potential V closely resembles a cosmological constant. In 
practice, the potential alone would not be able to induce any dynamics 
to the field.  
The energy density of the scalar field vs. redshift is plotted in 
Fig. \ref{Figura1} for four different initial conditions 
${\Omega_Q}_{init}$, spanning several orders of magnitude. 
In each case, the initial kinetic energy of the field has been taken to 
vanish, while we changed the initial value of the field in a range of 
6 orders of magnitude; independently of the initial conditions, the
Quintessence field is found to finish up with $\dot{w}_Q \sim 0$.
The freezing is reached later for higher 
${\Omega_Q}_{init}$ (see \cite{BMP}). 
In the same figure, we can distinguish the scaling of the matter and 
radiation components.  
For comparison, we also plotted in Fig. \ref{Figura2} the scaling of 
the DE component in a minimally-coupled model with the same inverse power-law 
potential, the same equation of state today and the same initial values 
of the field as in Fig. \ref{Figura1}. 
In the latter case, because of the potential flatness,  
very different initial field values span a narrow
range of initial energy densities, unlike Extended Quintessence; 
as discussed in \cite{Bludman}, the initial energy density has to be 
tuned in a tiny interval, here covering a range of a few tenths of 
${\rm meV}^4$ at $z \sim 10^8$. As $w_Q\rightarrow -1$, this basin of 
attraction shrinks to a single value, while in non-minimally 
coupling cases it remains huge, thereby  avoiding any 
fine-tuning of the initial energy density.  \\ 
From Fig. \ref{Figura1} we see how this loss of dependence on 
the initial conditions is realized in Extended Quintessence: 
the $R$-boost adds dynamics to the cosmological constant, which can now
be reached from a huge range of initial values. The only value which has 
to be fixed is the present DE energy density $\Omega_Q$. 
In conclusion, the $R$-boost, a purely gravitational effect 
characteristic of non-minimally coupled theories, has the attractive 
feature of enriching the set of initial conditions a field can have 
 to closely mimic the cosmological constant today.  
In canonical Quintessence scenarios, in order to reach $w_Q=-1$ 
today one would have to exactly tune the 
initial value $\Omega_Q$ to the present one; here, the balance between 
cosmological friction and gravitational coupling plays a fundamental 
role, enhancing the dynamics of the field down to relatively low 
redshifts and allowing to approach $w_Q=-1$ without fine-tuning on 
the initial energy density. \\
While, given the present experimental uncertainty on the value of the 
DE equation of state, canonical Quintessence models are still far 
from experiencing a real crisis, future precision measurements 
of $w_Q$ and  $\dot{w}_Q$ will be crucial: if experiments will 
converge towards $w_Q=-1$, 
and $\dot{w}_Q \rightarrow 0$, as it seems to be the case from the 
latest observations \cite{Riess2}, minimally-coupled Quintessence 
scenarios would suffer from a serious fine-tuning problem which would 
call for either an anthropic explanation or for new classes of models, 
among which Quintessence models based on scalar-tensor theories 
of gravity appear particularly promising.    

%
%
\bibliographystyle{}

\end{document}